\documentclass{emulateapj}

\usepackage{graphicx}
\usepackage{amsmath}
\usepackage{times}
\usepackage{natbib}
\newcommand{\kms}{\,{\rm km\,s^{-1}}}
\newcommand{\msun}{\,{\rm M_\odot}}
\newcommand{\beq}{\begin{equation}}
\newcommand{\eeq}{\end{equation}}
\def\lta{\mathrel{\rlap{\lower 3pt\hbox{$\sim$}}\raise 2.0pt\hbox{$<$}}}
\def\gta{\mathrel{\rlap{\lower 3pt\hbox{$\sim$}} \raise 2.0pt\hbox{$>$}}}
\def\simlt{\mathrel{\rlap{\lower 3pt\hbox{$\sim$}}\raise 2.0pt\hbox{$<$}}}
\def\simgt{\mathrel{\rlap{\lower 3pt\hbox{$\sim$}} \raise 2.0pt\hbox{$>$}}}
\begin{document}

\title{Sub-parsec supermassive Binary Quasars:  expectations at z$<1$}
\author{M. Volonteri\altaffilmark{1}, J. M. Miller\altaffilmark{1}, \& M. Dotti\altaffilmark{1}}
%\author{Michiganders\altaffilmark{1}}

%\email{martav@umich.edu}

\altaffiltext{1}{Department of Astronomy, University of Michigan, 500
Church Street, Ann Arbor, MI, USA}

\begin{abstract}
We investigate the theoretical expectations for detections of
supermassive binary black holes that can be identified as sub-parsec luminous quasars.  
To-date, only two candidates have been selected in a sample comprising 17,500 sources
selected from  the Sloan Digital Sky Survey (SDSS) Quasar Catalog at $z<0.70$ \citep{boroson09}. 
In this Letter, we use models of assembly and growth of supermassive black holes (SMBHs) in
hierarchical cosmologies to study the statistics and
observability of binary quasars at sub-parsec separations. Our goal is
twofold: (1) test if such a scarce number of binaries is consistent
with theoretical prediction of SMBH merger rates, and (2) provide
additional predictions at higher redshifts, and at 
lower flux levels.  We determine the cumulative number of expected binaries in a complete, volume limited sample.  
Motivated by \cite{boroson09}, we  apply the SDSS Quasar luminosity cut (M$_i<-22$) to our theoretical sample, deriving an upper limit to the observable binary fraction. We find that sub-parsec quasar binaries are intrinsically rare. 
Our best models predict $\sim 0.01$ deg$^{-2}$ sub-parsec binary quasars with separations below $\sim10^4$ Schwarzschild radii ($v_{\rm orb}> 2000 \kms$)   at $z<0.7$, which represent
a fraction $\sim 6\times10^{-4}$ of unabsorbed quasars in our theoretical sample. 
In a complete sample of $\sim$10,000 sources, we therefore predict an upper limit of  $\sim$10 sub-parsec 
binary quasars.  The number of binaries increases rapidly with increasing redshift. The decreasing lifetime with SMBH binary mass
suggests that lowering the luminosity threshold does not lead to a significant increase in 
the number of detectable sub-parsec binary quasars. 
\end{abstract}


\section{Introduction}
SMBHs appear to inhabit most galaxy centers
\citep[e.g.,][]{richstone1998, ferrareseford}, and in $\Lambda$CDM
cosmologies galaxies experience multiple mergers during their cosmic
assembly.  SMBH {\it binaries} (SMBHBs) are therefore expected to be
recurrent, albeit transient features of most galactic
bulges. Observationally, the paucity of quasar pairs ($\sim$0.1\%)
 on galactic scales \citep{foreman09} points toward rapid
inspiral of the SMBHs down to (sub-)parsec scales where they form a
gravitationally bound pair \citep{Mayer2007}.  
As discussed in \cite{foreman09} if we assume that every galaxy hosts a SMBH and that quasar activity is triggered by galaxy mergers, the probability of observing a double quasar scales with the ratio of the quasar lifetime to the total merger timescale as  $\simeq (t_{qso}/t_{merge})^2$, if  the two quasars light up at different, random times.  This is consistent with the fraction of quasar pairs found in the SDSS \citep{Hennawi}, if $t_{qso} \simlt 10^7 \mathrm{yr}$ and $t_{merge} \sim 10^{9} \mathrm{yr}$ \citep[for more details, see][]{foreman09}.
At lower level of activity, \cite{comerford09} find that about 2\% of early-type galaxies host candidate AGN pairs in the same galaxy. 

Detecting {\it sub-parsec binaries} by imaging techniques is extremely difficult.
However, the presence of SMBHBs in AGNs can be discovered spectroscopically, as
double broad--line emission systems (see \cite{Gaskell1996} and references
therein). The two sets of broad emission lines originate in gas
associated with the two SMBHs, and the velocity separation between the
two emission line systems traces the projected orbital velocity of the
binary. \cite{dotti09} and \cite{bogdanovic09} extended the pre--existent
spectroscopical technique, discovering that, depending on the SMBH
mass ratio, the AGN spectrum shows two sets of broad lines (equal mass
binaries) or a single set of broad lines and two sets of narrow
emission lines at different redshifts (unequal mass binaries).  In the
latter case, only one of the two SMBHs is active, and the two sets of
narrow emission lines correspond to emission from low density gas in
the potential well of the binary and from the ``standard'' narrow line
region (NLR) of the AGN.

\cite{dotti09} and \cite{bogdanovic09} apply the binary model to the
peculiar quasar SDSS J092712.65+294344.0. This quasar exhibits two
distinct sets of lines. The first set of very narrow emission lines is
assumed to be emitted in the NLR of the host and trace the redshift of
the host galaxy ($z=0.713$).  The second set comprises two
blue--shifted systems featuring different FWHM: the broad Mg II and
Balmer emission lines with FWHM $\approx 4000$ km s$^{-1}$, and narrow
lines with FWHM $\approx 460-2000$ km s$^{-1}$, both consistent with a
redshift of 0.698. In this model, the source emitting the
blue--shifted line system is gas inside or near the broad line region
(BLR) of the secondary, comoving with the SMBH with a light-of-sight
velocity of 2630 km s$^{-1}$ relative to the rest-frame of the host.

\cite{boroson09} developed a principal components analysis technique
that identifies sources having peculiar spectral characteristics.
They applied this procedure to the restframe optical spectrum of
$\sim$17,500 quasars.  Their sample comprises all quasars having
$z<0.7$ from the fifth release of the SDSS Quasar Catalog  \citep{Schneider2007} plus all sources classified as quasars in the seventh SDSS data release.  Of the 17,500
objects in their entire sample, only two objects have multiple
redshift systems consistent with the presence of a candidate SMBHB, SDSS
J092712.65+294344.0, and SDSS J153636.22+044127.0 \citep[but see Chornok et al. 2009;][]{Heckman2009,Shields,Wrobel,Gaskell}.

In this paper we assess the expected number of merging SMBHBs based on
realistic merger rates of SMBHs in hierarchical cosmologies, and
determine an upper limit to sub-parsec binary quasars detectable as double broad-line emission quasars 

\section{Black hole merger rate and quasar activity}
We trace the evolution of SMBHs within a plausible scenario for the
hierarchical assembly, growth, and dynamics of SMBHs in a $\Lambda$CDM
cosmology.  Our model has been shown to capture many features of the
SMBH population (e.g., luminosity function of quasars, X--ray
background, SMBH mass density). The main features of the models have
been discussed elsewhere \citep[and references
therein]{VHM,Volonterietal2005, VolonteriRees2006, gw3}. 
We summarize in the following the relevant assumptions. 

SMBH ``seeds" form at high redshift ($z>15$) in highly biased halos,
corresponding to 3.5--4 $\sigma$ peaks of the density
fluctuation field \citep{Volonterietal2007}. The initial SMBH occupation 
fraction is therefore low, but along the cosmic hierarchy 
SMBHs are incorporated into massive and massive systems,
as galaxies grow in a $\Lambda$CDM cosmology. 

The occupation fraction of SMBHs increases with time, and approaches unity 
for massive galaxies at low redshift \citep{Marulli2006,Volonterietal2007}. In our scheme, 
therefore, not all galaxy mergers lead to SMBH mergers, but only those involving 
two galaxies both hosting SMBHs. We further assume that SMBHs merge 
within the merger timescale of their hosts, which is a likely assumption for SMBH binaries
formed after gas rich galaxy mergers
\citep{Escalaetal2004,Escalaetal2005,Dottietal2006,Dotti2007}. 
This is the most likely scenario in the context of this work, as quasar fueling requires
a substantial gas supply. 
We explored an alternative scenario where at late cosmic times SMBHBs shrink via
three-body interactions, i.e., by capturing and ejecting at much
higher velocities the stars passing by within a distance comparable to
the binary separation \citep{Merritt2006,VHM, Sesana2007}, and we
found that the merger rate in the $z<1$ redshift range is very
similar, as already found in previous tests \citep{Sesanaetal2005}. 
\cite{Sesana08} compare our theoretical merger rates with the merger rates inferred by observations of the fraction of close galaxy pairs (assuming that SMBH masses scale with bulge masses).  Our SMBH merger rate is consistent Ð within a factor of  $\simlt$ 2 Ð with the merger rate of massive spheroidal found in Bell et al. (2006), who quote a factor of 2 uncertainty in their rate estimate. 

We base our model for SMBH mass growth on a set of simple assumptions, supported
by both simulations of AGN triggering and feedback
\citep{Springel2005b}, and analysis of the relationship between SMBH
masses ($M_{BH}$) and the properties of their hosts
\citep{Gebhardt2000,Ferrarese2000,Ferrarese2002}. 
SMBHs in galaxies undergoing a major
merger (i.e. having a mass ratio $>1:10$) undergo accretion. Each SMBH
accretes an amount of mass, $\Delta M=9\times 10^7\msun(\sigma/200\kms)^4$, that scales with the $M_{\rm
BH}-\sigma_*$ relation of its hosts (see Volonteri \& Natarajan 2009). Accretion starts after a
dynamical timescale and lasts until the SMBH has accreted $\Delta M$.

The accretion rate during the active phase is derived from the
empirical distribution of Eddington ratios, $\lambda=\log(L_{\rm
bol}/L_{\rm Edd})$, found in \cite{Merloni08}. We adopt a fitting
function of the Eddington ratio distribution as a function of SMBH
mass and redshift (Merloni 2009).  The distributions of Eddington ratios,
$f_\lambda$, are computed in 10 redshift intervals (from $z=0$ to
$z=5$) for 4 different mass bins ($6 < \log(M_{\rm BH}/\msun)< 7$, $7
< \log(M_{\rm BH}/\msun) < 8$, $8 < \log(M_{\rm BH}/\msun) < 9$, $9 <
\log(M_{\rm BH}/\msun) < 10$), and then fit with an analytic function
which is the sum of a Schechter function and a log-normal. The
Eddington ratio distributions are normalized so that at a given mass
and redshift, $\int_{\lambda_{min}}^ \infty
f_\lambda=1$. 
%$\lambda_{min}$ has to be computed numerically.
We adopt the bolometric corrections
presented in \cite{richards06}, and we correct for absorbed quasars 
according to model 4 in \cite{Lafranca2005}.

%\begin{deluxetable}{lcccccccc}
%%\tabletypesize{\footnotesize}
%\tablenum{1} \tablewidth{\columnwidth} \tablecaption{Minimum value
%required to normalize the Eddington ratio distribution for a
%representative set of SMBH masses and redshifts.}
%\tablehead{\colhead{~~~~~~~~~ $\log(M_{\rm BH}/\msun)$} & \colhead{z}
%& ~~~~~~~~~~~~$\lambda_{min}$} \startdata ~~~~~~~~~~~~ 7.0 & 0.0 &
%~~~~~~~~~~~~ -2.87 \\ ~~~~~~~~~~~~ 7.0 & 0.5 & ~~~~~~~~~~~~ -3.81 \\
%~~~~~~~~~~~~ 7.0 & 1.0 & ~~~~~~~~~~~~ -3.89 \\ ~~~~~~~~~~~~ 7.5 & 0.0
%& ~~~~~~~~~~~~ -3.06 \\ ~~~~~~~~~~~~ 7.5 & 0.5 & ~~~~~~~~~~~~ -4.04 \\
%~~~~~~~~~~~~ 7.5 & 1.0 & ~~~~~~~~~~~~ -4.12 \\ ~~~~~~~~~~~~ 8.0 & 0.0
%& ~~~~~~~~~~~~ -3.26 \\ ~~~~~~~~~~~~ 8.0 & 0.5 & ~~~~~~~~~~~~ -4.29 \\
%~~~~~~~~~~~~ 8.0 & 1.0 & ~~~~~~~~~~~~ -4.38 \\ ~~~~~~~~~~~~ 8.5 & 0.0
%& ~~~~~~~~~~~~ -3.49 \\ ~~~~~~~~~~~~ 8.5 & 0.5 & ~~~~~~~~~~~~ -4.55 \\
%~~~~~~~~~~~~ 8.5 & 1.0 & ~~~~~~~~~~~~ -4.64 \\ ~~~~~~~~~~~~ 9.0 & 0.0
%& ~~~~~~~~~~~~ -3.73 \\ ~~~~~~~~~~~~ 9.0 & 0.5 & ~~~~~~~~~~~~ -4.83 \\
%~~~~~~~~~~~~ 9.0 & 1.0 & ~~~~~~~~~~~~ -4.93 \\ ~~~~~~~~~~~~ 9.5 & 0.0
%& ~~~~~~~~~~~~ -3.99 \\ ~~~~~~~~~~~~ 9.5 & 0.5 & ~~~~~~~~~~~~ -5.12 \\
%~~~~~~~~~~~~ 9.5 & 1.0 & ~~~~~~~~~~~~ -5.23 \\ \enddata
%\end{deluxetable}

Our first check is on the cumulative number counts of quasars, regardless of binarity. 
We select all accreting SMBHs in our theoretical sample, and apply the same luminosity cut as in the SDSS
Quasar Catalog, M$_i<-22$.  This choice is motivated by the SDSS being the largest quasar 
catalog currently available, however we stress here that our models provide upper limits to the
number of observable binaries, in a complete volume limited sample. 
We compare our number counts to the expected number from the integration
of the bolometric luminosity function of quasars \citep{Hop_bol_2007} at M$_i<-22$. 
The comparison is shown in the top panels of Fig.~1 (top histogram and dashed curve). 

%Within the area covered by the SDSS, we find $\sim10^4$ quasars with
%M$_i<-22$ at $z<0.7$, in good agreement with the SDSS Quasar Catalog.

\section{Sub-parsec binary quasars}
To evaluate the expected number of binary quasars, we start from the
SMBHB merger rate. Our theoretical models indicate which of these
binaries are in an active phase, following a major merger. We
initially analyze the complete sample of SMBHBs, that is, we
consider all merging SMBHs, regardless of their activity status. This
first model (model I) provides a strong upper limit to the number of theoretical
sub-parsec binary quasars. To each SMBH in our sample we randomly
assign an Eddington ratio, $\lambda$, from the normalized 
distribution (Merloni 2009).
%For reference, in Table 1 we list the value of $\lambda_{min}$ for a representative set of SMBH masses and redshifts.
Note that, since $-5\simlt\lambda_{min}\simlt-3$ for most masses/redshift this model
does not allow for `quiescent' SMBHs at, e.g., the level of Sgr A* \citep{quataert1999}.

We further assume that only SMBHBs with a mass ratio, $q=M_{\rm BH,2}/M_{\rm BH,1}\leqslant1$, above a
certain threshold ($q>10^{-1}$ or $q>10^{-2}$) create distinguishable
double broad emission line systems. We regard $q>10^{-1}$ as our best choice, because of an additional 
independent motivation, as follows. \cite{callegarietal2009}  study the formation of  SMBHBs during galaxy mergers, 
and how SMBH pairing depends on  the interplay between different physical processes (dynamical friction, tidal and ram--pressure stripping). 
From this analysis, \cite{callegarietal2009}  find that  $q=10^{-1}$ is the minimum mass ratio between two merging galaxies 
(and as a consequence between the two SMBHs) to guarantee the 
formation of a SMBHB.  However, tidal and gas--pressure stripping can be 
reduced for extreme structural properties and orbital parameters of merging
galaxies. We consider the very conservative case of SMBHBs with
$q>10^{-2}$ to take into account every possible merger configuration 
(e.g., plunging radial orbits).

Finally, we assign a lifetime 
%(residence time at separations where the
%orbital velocity is a few thousand $\kms$, making the binary
%detectable through the line shifts) 
to SMBH binaries detectable at sub-parsec separations following the detailed study
of SMBHB dynamical evolution in circumbinary disks performed by
\cite{Haiman09}.  The lifetime corresponds to the time spent by the
SMBHB at a separation such that the shift between the BLRs (or the NLR
and the BLR) is a few thousand $\kms$, becoming comparable with the
widths of typical broad lines \citep{Shen2008}. Blending of profiles with
smaller velocity differences would be missed \citep{Gaskell1996}.
For $v_{\rm orb}> 2000 \kms$, the typical separation is $r=11.25 \times 10^3$ Schwarzschild radii, and for these conditions, \cite{Haiman09} give a lifetime:
\beq
t_{\rm life}=6 {\rm Myr} \left(\frac{M_{\rm bin}}{10^7 \msun}\right)^{3/4}\, \left(\frac{4\, q}{{1+q}^2}\right)^{3/8}\left(\frac{10^\lambda}{0.1}\right)^{-5/8},
\eeq
where $M_{\rm bin}=M_{\rm BH,2}+M_{\rm BH,1}$, and we have chosen the longest residence time, thus providing an upper limit to SMBHBs lifetime. 

\begin{figure}[thb]
\includegraphics[width=\columnwidth]{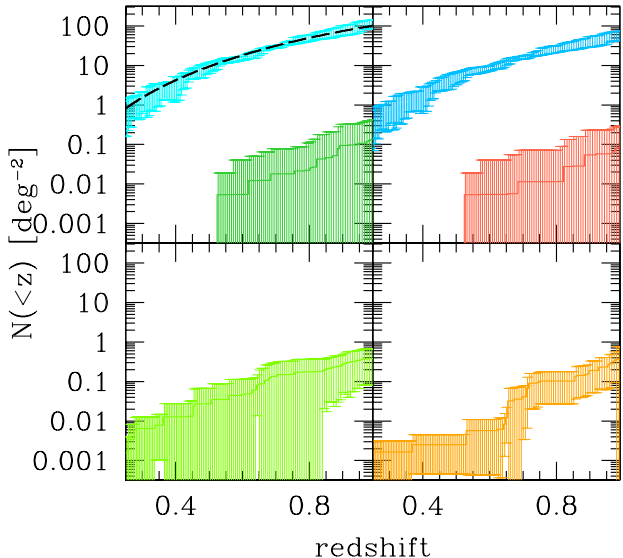}
\caption{Cumulative number of sub-parsec binary quasars with M$_i<-22$, per square degree. 
Left panels: $q>10^{-2}$. Right panels: $q>10^{-1}$. 
Bottom: we assume that each SMBH is at some level ``active",
i.e. we do not consider a merger-driven quasar activity (model I). Top: we
select only SMBHBs in galaxies that have recently experienced a major
merger, i.e. we do assume that quasar activity is merger driven (model II). 
%We require that at least one of the SMBHs in the pair is active. 
In the upper-left panel we compare our total quasar sample (top curve with error-bars) to 
the expected number counts from the integration of the bolometric 
luminosity function of quasars \citep{Hop_bol_2007} at M$_i<-22$ (dashed curve). 
In the upper-right panel we also show the number of unabsorbed quasars 
when we correct for absorption (top curve with error-bars).
%Right: we require that both SMBHs in the pair are still active after the
%merger, but only one needs to be above the SDSS quasar luminosity
%threshold (model III). 
}
\label{fig1}
\end{figure}

The results of this model (model I) are shown in Fig.~\ref{fig1} (bottom panels). We find that at
$z<0.7$ the total expected number of binaries with M$_i<-22$ per deg$^2$ is $0.27\pm0.27$ for $q>10^{-2}$
and $0.04^{+0.06}_{-0.04}$ for $q>10^{-1}$. These detectable binaries represent a fraction $\sim10^{-2}$ and 
$\sim2\times10^{-3}$ respectively of the unabsorbed quasars. 
%Since the SDSS quasar catalog is not a complete sample \citep{Schneider2007}, we will base our estimate the upper limit to the number of binaries that can be found in the SDSS quasar catalog using this fraction. 
Within the sample of 17,500 sources analyzed by \cite{boroson09}, 9895 objects, including the two putative binary SMBHs, belong to the uniformly selected statistical sample of SDSS quasars (Boroson, private communication). The statistical sample reaches a completeness larger than 90\% at $z<1$ for sources with apparent magnitude $i<19.1$, roughly corresponding to M$_i\simlt 24$ at $z<0.7$ (Richards et al. 2002). Given these caveats, our results must be considered upper limits to the number of detectable sub-parsec binary quasars.  We therefore find marginal agreement, within the uncertainties,  with \cite{boroson09} findings\footnote{We have calculated the error by using the simple Poisson statistics, 1--$\sigma$ confidence. The errors are {\it lower limits} given the small number statistics. See \cite{gehrels86}.}.  If binaries with $q \lta 10^{-2}$ do produce distinguishable double broad emission line systems, then the expected SMBHB merger rate must be lower than we predict. 
%Fig.~2 shows the mass and Eddington ratio distributions for the sources, including all primary SMBHs if they shine as quasars above threshold (i.e., M$_i<-22$).  
The high luminosity selection criterion leads to a sample composed of actively accreting ($\lambda \sim -1$) massive ($\sim10^8 \,\msun$) SMBHs.

%\begin{figure}[thb]
%\includegraphics[width=\columnwidth]{f2_col.eps}
%\caption{Number of sub-parsec binary quasars per square degree,
%if the luminosity threshold is decreased by one order of magnitude, i.e. $L_{bol}>10^{44}\,{\rm erg\,s}^{-1}$.
%Left: we assume that each SMBH is at some level ``active",
%i.e. we do not consider a merger-driven quasar activity (model I).
%Right: we require that at least one of the SMBHs in the pair is active,
%following a major merger (model II). Top panels: $q>10^{-2}$. Bottom
%panels: $q>10^{-1}$. }
%\label{fig2}
%\end{figure}

In a second model (model II)  we select only SMBHBs where at least one of the 
SMBHs is active, according to our merger-driven quasar activity scheme.  
%We again set the lifetime to $10^7$ years and investigate binaries with $q>10^{-2}$ or
%$q>10^{-1}$. We then select quasars with M$_i<-22$. 
The results of this model are shown in the top panels of Fig.~1. In this case the expected
number of binaries per deg$^2$  is $0.02^{+0.06}_{-0.02}$ for $q>10^{-2}$ and $0.01^{+0.05}_{-0.01}$ for
$q>10^{-1}$. Detectable binaries represent a fraction $\simlt 6\times10^{-4}-10^{-3}$ of the unabsorbed quasars, consistent with \cite{boroson09} findings (a fraction $\simeq 1-2\times 10^{-4}$). 
%Note, if we assumed that the lifetime is $10^6$ years (as would be appropriate for binaries with total mass $10^7-10^8\, \msun$, Haiman et al. 2009), the fraction of detectable binaries decreases to $\sim 6\times10^{-5}$ for $q>10^{-1}$, and $\sim 4\times10^{-5}$ for $q>10^{-2}$.

%\begin{figure}[thb]
%\includegraphics[width=\columnwidth]{f2.eps}
%\caption{Distribution of primary SMBH masses (bottom panel) and Eddington
%ratios (top panel) of detectable sub-parsec binaries for model I.}
%\label{fig3}
%\end{figure}

\begin{figure}[thb]
\includegraphics[width=\columnwidth]{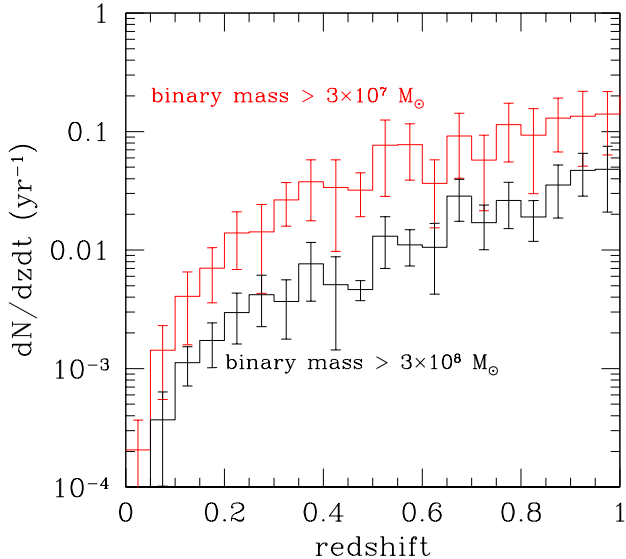}
\caption{Merger rate of SMBHs per unit redshift. Top histogram: binary mass $>3\times 10^7\, \msun$. Bottom histogram: binary mass $>3\times 10^8\, \msun$.}
\label{fig2}
\end{figure}

If we decrease the luminosity threshold we expect two factors enter into play: on the one hand, the merger rate of SMBHs is expected to increase at lower SMBH masses, where the mass function is less  steep \citep{Gultekin2009}. This is indeed what we find in our models (Fig.~\ref{fig2}; see also Dotti et al. 2009 and Sesana et al. 2005). On the other hand, however, the lifetime of detectable binaries decreases with decreasing mass  \citep{Haiman09}, making the detection harder. Using the scaling for lifetimes presented in Haiman et al. 2009, we indeed expect that the number of detectable sub-parsec binary quasars does not increase dramatically with decreasing luminosity, because of the shorter timescale over which they are observable. For instance, if we decrease the flux limit by a factor of ten, we find negligible changes in model II, and a mild increase in the number of binaries by a factor of 3 in model (I): $0.17\pm 0.13$ for $q>10^{-1}$ and $0.79\pm 0.64$ for $q>10^{-2}$. %If we decrease the flux limit by a factor 100, model I number counts increase, again, very mildly to $0.06^{+0.30}_{-0.04}$ for $q>10^{-2}$ and $0.04^{+0.29}_{-0.04}$ for $q>10^{-1}$, and model II increases to $0.65\pm 0.44$ for $q>10^{-1}$ and $1.91\pm 1.40$ for $q>10^{-2}$. The binaries that cause the increase in model I are composed of massive black holes (>10$^8\, \msun$) accreting at low Eddington ratios. 

\section{Discussion}
Stimulated by the recent putative discovery of two candidate sub-parsec SMBHBs identified 
as quasars with multiple redshift line systems (BLR and NRL), SDSS J092712.65+294344.0 and 
SDSS J153636.22+044127.0, we investigate theoretically the occurrence of sub-parsec 
SMBHBs that can be identified as sub-parsec binary quasars.
We study the SMBH cosmic evolution via a Monte-Carlo merger tree approach. 
We trace the growth and dynamical history of SMBHs from high redshift via physically motivated
prescriptions, that allow us to reproduce many observational constraints.

Our approach provides us with a catalog of SMBHBs, for which we know the masses and the redshift.
We further assume that only SMBHBs with a mass ratio,  $q=M_{\rm BH,2}/M_{\rm BH,1}\leqslant1$, above a certain threshold 
($q>10^{-1}$ or $q>10^{-2}$) create distinguishable double broad emission line systems.
Motivated by the work by Boroson \& Lauer (2009), we apply to our theoretical sample the same luminosity cut as in the SDSS Quasar Catalog, M$_i<-22$,
thus deriving upper-limits to the fraction of detectable binaries. We stress here that the theoretical sample is complete and volume limited. Since the SDSS quasar catalog is not a complete sample \citep{Schneider2007}, a direct comparison with Boroson \& Lauer (2009) is not appropriate. However, our results are consistent with the binary fraction derived for the subset of the 17,500 quasars used by Boroson \& Lauer that are part of the statistical sample of the SDSS. 

Our merger-driven quasar scheme provides us also with an accretion rate, hence a luminosity.  We
analyze two models that likely bracket the theoretical uncertainties. Model I ignores merger-driven
quasar activity and we assume that each SMBH is at some level ``active".  Each SMBH in our binary  sample
is assigned an Eddington ratio, $\lambda$, from the normalized  distribution derived from synthesis 
model for AGN evolution \citep{Merloni08}. Model I is therefore our strong upper limit to the number of detectable SMBHBs. 
Model II is more rooted into our quasar activity scheme, as we select SMBHBs where at least one of the 
SMBHs is active, according to our merger-driven scenario. 
%Finally, we implemented a very conservative model
%where we require that both SMBHs in the pair are still active after the merger, but only one needs to be above
%the SDSS quasar luminosity threshold. 

Our main findings are as follows:

\begin{itemize}
\item Sub-parsec binary quasars are intrinsically rare, due to a combination of strict requirements: the time over which SMBHB are detectable through line shifts decreases with decreasing binary mass. On the other hand, the merger rate of SMBHBs increases with increasing mass. 
\item   In a volume limited, complete sample of $\sim$ 10,000 sources at $z<0.7$, our best models (II), that relate quasar activity to galaxy mergers,  predict an upper limit of  $\sim$ 5-10 sub-parsec binary quasars. Model I, that does not associate quasars to mergers, is only marginally compatible with  \cite{boroson09} who find only 2 candidate sub-parsec binary quasars in the statistical SDSS quasar sample. 
% Our best models are consistent with the observations within the uncertainties. 
%\item If the lifetime of sub-parsec SMBHBs is short,  e.g., $10^6$ years, the number of detectable binaries is highly suppressed, as we would expect $1\pm1$ binaries with $q>10^{-2}$ (both model I and model II).
\item Fig.~1 extends our predictions out to $z=1$. The number of detectable binaries increases by a factor $\sim 5-10$ from $z=0.7$ to $z=1$.
%\item The selection criteria adopted for SDSS quasars selects high-mass ($10^8-10^9\,\msun$)  SMBHs that typically accrete efficiently.   
\item The lifetime over which SMBHBs can be detected as sub-parsec quasars decreases with decreasing binary mass  \citep{Haiman09}.  This effect is stronger than the increase in the merger rate of SMBHB at lower masses. Lowering the luminosity threshold is unlikely to lead to a large increase in the number of detectable sub-parsec binary quasars. 
\end{itemize}

SDSS-III will increase the spectroscopic quasar sample and will provide a good testing ground for our predictions. The masses of SMBHBs that can be identified as sub-parsec binary quasars are too large for the gravitational waves emitted by these binaries to be detectable at merger by the {\it Laser Interferometer Space Antenna} ({\it LISA}), which will instead focus on the mass range $10^5-10^7\msun$. However, such massive binaries (in a later evolutionary stage, when the binary has shrunk by an additional factor of ten and the dynamical evolution is driven by emission of gravitational radiation) are typical candidates for detection via Pulsar Timing Arrays (PTAs, e.g. the Parkes radio-telescope). PTAs rely on the effect of gravitational waves on the propagation of radio signals from a pulsar to the Earth, producing a characteristic signature in the time of arrival of radio pulses. \cite{PTA} find that the mass distribution of the SMBHBs detectable via PTAs peaks at $\sim10^8\,\msun$, with most binaries at $z<1$, the same mass range probed by sub-parsec binary quasars identifiable in the SDSS. \\

\acknowledgements We thank Tim McKay and Douglas Tucker for help with
the SDSS magnitude system. We also thank Armin Rest for suggestions on
the statistical treatment of data. We gratefully acknowledge help from
Todd Boroson, Tod Lauer and Gordon Richards for the comparison between models and data. 

%\bibliographystyle{apj}
%\bibliography{../marta}
%\end{document}

\end{document}